\documentclass[useAMS,usenatbib]{mn2e}
\usepackage{graphicx}
\usepackage{amsmath}

\footnotesize
\newdimen\digitwidth    
\setbox0=\hbox{\rm0}
\digitwidth=\wd0
\catcode`!=\active
\def!{\kern\digitwidth}
\normalsize

\title[The mass of PSR\,J1738+0333]{The relativistic pulsar-white dwarf binary PSR\,J1738+0333
\\ I. Mass determination and evolutionary history}

\author[J. Antoniadis et al.]{J.~Antoniadis,$^1$
\thanks{Member of the International Max Planck Research School (IMPRS)
for Astronomy and Astrophysics at the Universities of Bonn and Cologne}
\thanks{e-mail: jantoniadis@mpifr-bonn.mpg.de}
M.~H.~van Kerkwijk,$^2$ D.~Koester,$^3$ P.~C.~C.~Freire,$^1$ N.~Wex,$^1$ 
\newauthor T.~M.~Tauris,$^{1,4}$, M.~Kramer,$^{1,5}$ and C.~G.~Bassa$^5$
\\
$^{1}$Max-Planck-Institut f\"{u}r Radioastronomie, Auf dem H\"{u}gel 69, 53121 Bonn, Germany
\\
$^{2}$Department of Astronomy and Astrophysics, University of Toronto, 50 St. George Street, Toronto, ON~M5S~3H4, Canada
\\
$^{3}$Institut f\"{u}r Theoretische Physik und Astrophysik, Universit\"{a}t Kiel, Kiel, 24098, Germany
\\
$^{4}$Argelander Institut f\"{u}r Astronomie, Auf dem H\"{u}gel 71, 53121 Bonn, Germany
\\
$^{5}$Jodrell Bank Centre for Astrophysics, The University of Manchester, 
Alan Turing Building, Manchester, M13 9PL, UK }
\date{Received: --Accepted:}
\begin{document}

\maketitle

\newcommand{\setthebls}{

}

\begin{abstract}
 PSR J1738+0333 is one of the four millisecond pulsars known to
 be orbited by a white dwarf companion bright enough for optical
 spectroscopy.  Of these, it has the shortest orbital period, making
 it especially interesting for a range of astrophysical and gravity
 related questions.  We present a spectroscopic and
 photometric study of the white dwarf companion and infer its radial velocity
 curve, effective temperature, surface gravity and luminosity. 
  We find that the white dwarf has properties consistent with those of low-mass white
 dwarfs with thick hydrogen envelopes, and use the corresponding
 mass-radius relation to infer its mass; $M_{\rm WD} =
 0.181^{+0.007}_{-0.005}$\,M$_{\odot}$.  Combined with the mass ratio
 $q=8.1\pm0.2$ inferred from the radial velocities and the precise pulsar
 timing ephemeris, the neutron star mass is constrained to $M_{\rm PSR} =
 1.47^{+0.07}_{-0.06}$\,M$_{\odot}$.  Contrary to expectations, the
 latter is only slightly above the Chandrasekhar limit. We
 find that, even if the birth mass of the neutron star was only 1.20\,M$_\odot$,
  more than 60\% of the matter
 that left the surface of the white dwarf progenitor escaped the
 system.
 The accurate determination of the component masses transforms this
 system in a laboratory for fundamental physics by constraining the orbital decay predicted by general
 relativity.  Currently, the agreement is within $1\sigma$ of the observed
 decay. Further radio timing observations will allow precise
 tests of white dwarf models, assuming the validity of general relativity.

\end{abstract}

\section{Introduction}
Millisecond pulsars (MSPs) are extreme in many ways. Their interior
consists of the densest form of observable matter known and
they can spin at least as fast as 716 times per second \citep{hrs+06}. 
Hence, they offer a rich laboratory for a wide range of
physical inquiry: Mass measurements provide direct comparison to
quantum chromodynamics' predictions for the state of ultra-dense
matter \citep{lp04,dpr+10} and studies of their orbits in binaries
have provided the first confirmation for gravitational wave emission
and the most stringent strong-field tests of general relativity
\citep{tw82,wnt10,ksm+06}.  

Most of the fastest spinning Galactic-disk pulsars are
paired with low mass helium-core WDs \citep[hereafter LMWDs, for 
recent reviews see][]{lrr,t11}, and their fast spins and weak magnetic
fields are thought to be the product of mass transfer from the
progenitor of the WD, a process also known as recycling. As the
progenitor star evolves, it fills its Roche lobe and loses its
envelope, either while on the main sequence (for sufficiently short
initial periods), or when moving up the red-giant track \citep{wrs83}.
The mass transfer rate is a strongly increasing function of the
initial orbital period and donor mass \citep{ts99a}, and is expected
to be at a stable, sub-Eddington rate
($\la\!10^{-8}$\,M$_{\odot}$\,yr$^{-1}$) for light companions in
relatively tight orbits. The final result of such long-term (nuclear
timescale) mass transfer is a highly circular (due to fast tidal
dissipation in the secondary) close binary consisting of a fast spinning
 MSP and a low mass, helium-core WD.

These systems are important for several reasons. First, it is these binaries
 that allow one to probe certain aspects of the radiative properties of gravity
 that are poorly constrained by the relativistic effects seen in double neutron stars, 
like the Hulse--Taylor or the double pulsar. 
For example, in a wide range of theories, the rate of gravitational
 wave emission is driven by a leading dipolar term that depends crucially
on the difference in gravitational binding energies between the binary members. 
Hence, if accurate component masses can be determined,
one can directly confront the predictions of different gravity theories in terms
of dipolar radiation with observations.

Second, measuring their masses provides access to the accretion process and
evolution of these systems as well as the formation of MSPs, the only neutron stars with secure precise masses significantly above
the Chandrasekhar limit \citep{fbw+11,dpr+10}. 
In addition, observational constraints on the upper mass limit of
stable neutron stars, constrains the equation of state for super-dense
matter.

Unfortunately, precise MSP and companion masses can be determined from
timing in exceptional cases only: either when the orbit is
(unexpectedly) eccentric, allowing for a measurement of the rate of
advance of periastron \citep{fbw+11}, or if the system has an orbit seen edge on
\citep{ktr94,jhb+05,dpr+10} which allows for
a measurement of pulse time-of-arrival (TOA) delays due to the
curvature of space-time around the companion \citep[Shapiro delay,][]{sha64}.  

Fortunately, another method exists that relies on combined optical and radio timing
observations \citep{vbk96,cgk98}.  If the WD companion is
bright enough for detailed spectroscopy,
a comparison of its spectrum with model
atmospheres yields its effective temperature and surface
gravity. These can then be compared with a mass--radius relation for
LMWDs to yield its mass.  Combining the radial velocity for the white
dwarf with the pulsar timing measurements yields the mass ratio and
therefore the mass of the pulsar.  

In a companion paper (Van Kerkwijk et al.\ 2011, in prep.; vK+12
hereafter), we test this method on PSR J1909$-$3744, for which the
masses are precisely known from timing.  We find it reliable and are confident to apply it also to other similar systems.
Here, we report on an application to PSR\,J1738+0333, a pulsar-LMWD
binary in a 8.5\,hours orbit.  This system is particularly interesting
because of its short period, which places it in a poorly studied
regime for accretion physics of neutron stars, where nuclear-driven evolution competes
with magnetic braking and gravitational radiation
\citep[c.f.][]{pk94}.  Furthermore, the short orbital period implies
relatively rapid orbital decay, making the system interesting for
radiative tests of gravity, for which, as described above, knowledge
of the component masses is necessary.

The paper is organized as follows: In \S\,\ref{section:2} we report
results from radio timing, necessary for calculations throughout the
rest of the paper.  These are presented in detail separately
\citep[][from now on Paper\,II]{fwe+12}. In \S\,\ref{section:3}
we describe the spectroscopic and photometric observations and in
\S\,\ref{section:4} we present our results. We discuss our findings
and comment on the evolution of the system and its importance for
gravity tests in \S\,\ref{section:5}. Finally, in \S\,\ref{section:6},
we summarize our results.

\section{PSR\,J1738+0333: Radio Observations}\label{section:2}
PSR\,J1738+0333 was discovered in a 20-cm high Galactic
latitude survey in 2001 \citep{jac05}, carried out with the
multi-beam receiver of the Parkes Telescope.  The pulsar has a spin period of
5.85\,ms and orbits a low-mass helium-core WD companion in a 8.5\,h
orbit.  Since 2003 it has been regularly timed with the 305\,m Arecibo
Telescope, leading to $\sim\!17000$ times
of arrival with typically 3\,$\rm{\mu}$s uncertainties.  The
corresponding timing solution provides
measurements of the system's parallax and proper motion, and
a significant detection of the intrinsic orbital period derivative (see
Paper\,II for details).  In Table\,\ref{table:2} we list the measured spin,
Keplerian and astrometric parameters of the system.

The spin period derivative is that of a typical low-surface magnetic
field pulsar ($B_{0} = 3.7 \times 10^8$\,G), and the characteristic age
($\equiv P/2\dot{P}$) after subtracting the kinematic effects
(Paper\,II) is 4.1\,Gyr.  The parallax measurement corresponds to a
distance of $d=1.47\pm 0.10$ kpc.  The system's proper motion combined
with the parallax implies transverse velocities of $v_{\alpha} = d
\mu_{\alpha} = 49$\,km\,s$^{-1}$ and $v_{\delta} = d
\mu_{\delta}=36$\,km\,s$^{-1}$ in $\alpha$ and $\delta$
respectively. In \S\ref{section:4} we combine these values with the
systemic radial velocity, $\gamma$, to derive the 3D spatial velocity
and calculate the Galactic orbit of the binary.  The estimate for the
orbital eccentricity is one of the lowest observed in any binary system: When Shapiro delay is accounted for in the solution (Paper\,II), the
apparent eccentricity diminishes to $e = (3.5 \pm 1.1) \times
10^{-7}$.  We discuss the implication of this for evolutionary
scenarios in \S\ref{section:5}.
\section{Optical observations}\label{section:3}
\subsection{Spectroscopy}
Our main data set consists of eighteen long-slit phase resolved
spectra of PSR\,J1738+0333, obtained with the Gemini South telescope
at Cerro Pach\'on on ten different nights between April and June 2006
(see Table\,\ref{table:1}). For our observations we used the Gemini
Multi-Object Spectrograph (GMOS-S). The GMOS detector consists of
three $2048\times 4608$ EEV CCDs, each of which was read-out at
$2\times 2$ binning by a different amplifier, giving a scale of
0\farcs14 per binned pixel in the spatial direction, and, with the
1200\,lines\,per\,mm B1200 grism, 0.4\,\AA\ per binned pixel in the
dispersion direction.  We chose a relatively wide, 1\farcs5 slit, to
minimize atmospheric dispersion losses (see below).  This meant that
the resolution was set by the seeing, at $\sim\!3\,$\AA, or
$\sim\!200{\rm\,km\,s^{-1}}$ at 4300\,\AA.  In order to cover the
higher Balmer lines, we centred the grating at 4300\,\AA, for a
wavelength coverage from 3500 to 5100\,\AA.

\begin{table*}
\begin{center}
\caption{Log of observations and radial velocity measurements.}
\label{table:1}
\begin{tabular}{@{}cccccccccc}
\hline
       &                        &               & $v_{\rm R}$& $v_{\rm{WD}}$& $\Delta v$ &\\
Date  &   MJD$_{\rm{mid,bar}}$  & $\phi$   & (km\,s$^{-1}$) & (km\,s$^{-1}$) & (km\,s$^{-1}$) & $\Delta B'$\\
        & (1) & (2) & (3) & (4) & (5) & (6) \\
\hline
\textbf{Gemini,GMOS-S} \\

      2006\,Apr\,27 & 53852.310219 & 0.0250  &  $+56.3 \pm 0.8$ & $-209 \pm 27$ & $-265\pm 27$ & $2.91 \pm 0.06 $\\
                    & 53852.366314 & 0.1831  &  $+49.1 \pm 0.8$ & $-143 \pm 26$ & $-192\pm 26$ &$2.88 \pm 0.06 $ \\
      2006\,Apr\,28 & 53853.295453 & 0.8019  &  $+53.7 \pm 0.5$ & $-100 \pm 14$ & $-154 \pm 14$& $2.88 \pm 0.04 $ \\
                    & 53853.350638 & 0.9575  &  $+68.8 \pm 0.6$ & $-185 \pm 15$ & $-254 \pm15$ &$2.88 \pm 0.04 $ \\
      2006\,May\,07 & 53862.333037 & 0.2749  &  $+40.1 \pm 0.5$ & $ -35 \pm 13$ & $-75 \pm 13$ & $2.83 \pm 0.03 $ \\
                    & 53862.391933 & 0.4409  &  $+84.9 \pm 0.6$ & $+162 \pm 21$ & $+77 \pm21$ &$2.87 \pm 0.04 $ \\
      2006\,May\,26 & 53881.198674 & 0.4489  &  $+60.8 \pm 1.1$ & $+121 \pm 37$ & $+60 \pm 37$ & $3.03 \pm 0.11 $ \\
                    & 53881.252933 & 0.6018  &  $+67.0 \pm 0.9$ & $ +33 \pm 32$ & $-34 \pm 32$ &$2.97 \pm 0.09 $ \\
      2006\,May\,27 & 53882.352291 & 0.7005  &  $+43.1 \pm 0.5$ & $  -6 \pm 15$ & $-49 \pm 15$& $2.85 \pm 0.04 $ \\
      2006\,May\,28 & 53883.296760 & 0.3625  &  $+54.9 \pm 0.6$ & $ +28 \pm 16$ & $-27 \pm 16$  &$2.86 \pm 0.03 $ \\
                    & 53883.350144 & 0.5130  &  $+53.9 \pm 0.6$ & $+134 \pm 17$ & $-189 \pm 17$ &$2.85 \pm 0.03 $\\
     2006\,Jun\,19  & 53905.174549 & 0.0264  &  $+41.5 \pm 0.5$ & $-226 \pm 12$ & $+80 \pm12 $ &$2.85 \pm 0.04 $ \\
     2006\,Jun\,23  & 53909.170100 & 0.2881  &  $+39.9 \pm 0.7$ & $  -5 \pm 14$ & $-45 \pm 14$ &$2.88 \pm 0.03 $ \\
                    & 53909.210618 & 0.4023  &  $+95.6 \pm 1.1$ & $-103 \pm 36$ & $-45 \pm 36$ &$2.84 \pm 0.04 $ \\
     2006\,Jun\,26  & 53912.156147 & 0.7045  &  $+60.4 \pm 0.5$ & $  -7 \pm 14$& $-67 \pm 14$& $2.85 \pm 0.04 $ \\
                    & 53912.209838 & 0.8558  &  $+60.7 \pm 0.6$ & $-136 \pm 15$ & $-197 \pm15$ &$2.88 \pm 0.04 $\\
     2006\,Jun\,27  & 53913.120000 & 0.4212  &  $+42.5 \pm 0.5$ & $ +79 \pm 12$& $+37 \pm 12$& $2.84 \pm 0.04 $\\
                    & 53913.176660 & 0.5809  &  $+43.6 \pm 0.5$ & $+106 \pm 13$ & $+62\pm 13$ &$2.82 \pm 0.04 $ \\
\textbf{Keck, LRIS} \\ 
     2008\,Aug\,04 & 54682.377697 & 0.6224 & $ 50 \pm 1, +61 \pm 5 $ &$-2 \pm 9$ & $-52 \pm 9 $& $3.01 \pm 0.05$\\
\hline

\end{tabular}
\\
Notes: (1) refers to the barycentric mid-exposure time. (2) is the orbital phase using the ephemeris in Table\,\ref{table:2}.
(3) is the comparison's velocity in respect to the solar system barycenter and (4) the raw barycentric velocities of
PSR\,J1738+0333. (5) is the differential velocity used to determine the orbit in \S\ref{section:4}. Finally, (6) are the differential spectrophotometric
magnitudes in $B'$ (equal to $B$, but limited to the
wavelength range covered by our spectra; see \S\ref{section:3}).
Here, the errors are the quadratic sum of the photometric uncertainties of the WD and the comparison. 
For LRIS, two velocities are listed for the comparison star, for the
blue and red arm, respectively. For the white dwarf, the velocity is for
the blue arm (see text).
\end{center}
\end{table*}

\begin{figure}
\includegraphics[width=\hsize]{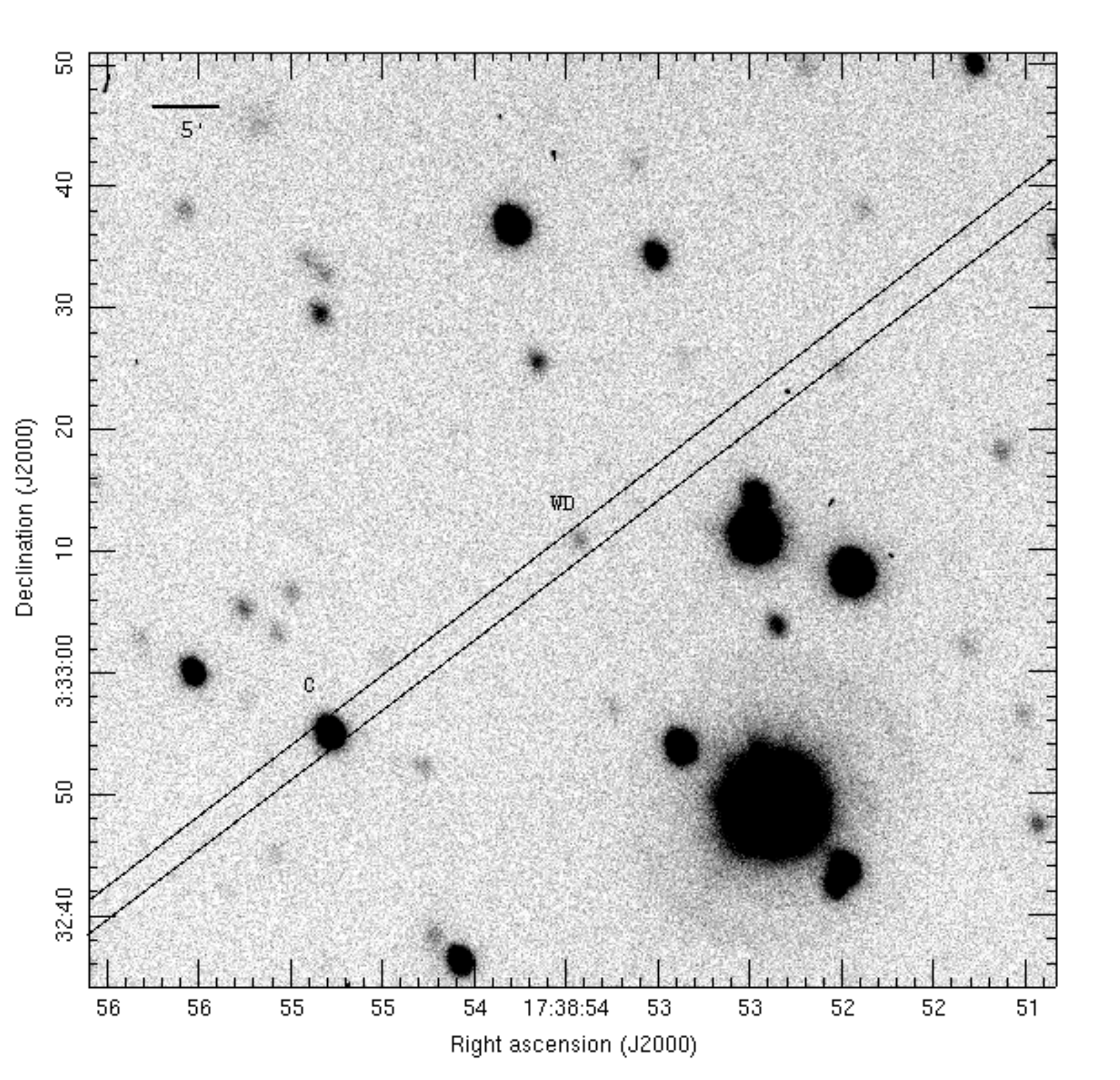}
\caption{Finding chart for PSR\,J1738+0333 (using the SOAR $V$ image).
 Indicated are the white-dwarf counterpart, the slit orientation
 used, and the comparison star that was included in the slit.}
\label{figure:1}
\end{figure}

All exposures had integration times of 3720\,s and were followed by an
internal flat-field exposure and a Copper-Argon (CuAr)
exposure for wavelength calibration. The slit was oriented to
include a bright comparison star located $25\farcs2$ at position angle
127\fdg57 (north through east) of the WD
(see Fig.~\ref{figure:1}).  We use this star as a local velocity
and flux standard (since GMOS-S does not have an atmospheric
dispersion corrector, slit losses vary with offset from the
parallactic angle).  

The conditions during the observations were mostly good to
photometric, but some exposures were taken through thin cirrus.  The
seeing ranged from $0\farcs6$ to $1\farcs2$.  For flux calibration,
we acquired additional frames of the comparison star and the
spectro-photometric standard EG\,274 through a $5\farcs0$ slit on the
night of 2006 April 27 (which was photometric and had $0\farcs8$
seeing).  Furthermore, for absolute velocity calibration, we observed
the radial velocity standard WD~1743$-$132 on 2006 June 19.

The data were reduced using standard and custom routines inside the
Munich Image and Data Analysis System (\textsc{midas}).  First, the
bias level of each exposure was removed using average values from the
overscan region.  Subsequently, we corrected the raw counts on the red
and middle chips for the small, few percent variations in gain (see vK+12 for details on the method), that
affected several sets of exposures (but fortunately not those of the
night the flux calibrator was taken).  Finally,
the frames were corrected for small-scale sensitivity variations using
normalised lamp exposures, where the normalisation was done both along
each wavelength position as well as along each spatial position.
These normalisation steps were required since the lamp spectra showed
rather sharp bumps in the dispersion direction whose position and
shape was different from bumps seen in target spectra, and also varied
between sets of spectra (possibly because the holographic grating was
not illuminated exactly identically between the different exposures),
while in the spatial direction they showed striations due to
irregularities in the slit. 

For sky subtraction, we selected a $100\arcsec$ region centred on the
WD, but excluding $5\arcsec$ spots around it and the comparison star.
Each column in the spatial direction was fitted with a second degree
polynomial and the interpolated sky contributions at the positions of
the WD and the comparison were removed.

Optimally weighted spectra and their uncertainties were extracted 
 using a method similar to that of \citet{h86}. The
extraction was done separately in each chip and the spectra were
merged after flux calibration.

The dispersion solution was established using the CuAr spectra taken
after each exposure.  First, the 1D lamp spectrum was extracted by
averaging the signal over the spatial direction in areas of the chip
that coincided with  each star.  Then the lines'
positions were measured and identified and the dispersion relation was
approximated with a $3^{\rm rd}$ degree polynomial that gave root-mean-square
residuals of less than 0.04\,\AA\ for typically 18 lines.

The wide-slit spectra of EG\,274 and the comparison star were
extracted with the same procedure and used to calibrate the
narrow-slit exposures.  Initially, all wide and narrow slit data were
corrected for atmospheric extinction using the average extinction
table for La Silla (which should be a good approximation to that of
Cerro Pach\'on).  Then, we calculated the wavelength-dependent flux
losses due to the finite size of the slit by comparing the wide-slit
spectrum of the comparison with each of the narrow-slit spectra.  The
relation was analytically approximated with a quadratic function of
wavelength that was then applied to the narrow slit
observations. Finally, the GMOS instrumental response was calculated
by dividing the spectrum of EG\,274 with a synthetic template and
smoothly interpolating the ratio.  The template was created by
normalizing an appropriate DA model atmosphere to the catalogued flux
\citep[$V = 11.03$,][see \S4.5 for more details on the model
atmospheres used in this work]{zcm04}.  Prior to comparison, we
smoothed the template with a Gaussian kernel to match the resolution
of the observed spectrum and excluded the cores of the Balmer lines.

Given the possible issues with the detector gain and the flat
fielding, both of which could affect the flux calibration, we obtained
an additional smaller set of spectra of the WD companion, the
comparison star and the spectro-photometric standard Feige\,110 using
the two-armed Low Resolution Imaging Spectrometer \citep[LRIS][]{lris}
of the Keck telescope on the night of 2008 August 3
(Table~\ref{table:1}).  During the night the sky was photometric and
the seeing was $\sim\!0\farcs8$.

For the observations we used the atmospheric dispersion corrector and
both narrow, $0\farcs7$, and wide, $8\farcs7$ slits.  The light was
split with a dichroic at 6800\,\AA\ and directed on the two arms of
LRIS (blue and red arm hereafter).  On the blue arm we used a
600\,lines\,mm$^{-1}$ grism, blazed at 4000\,\AA, that covers
3100--5600\,\AA\ with a resolution of $\Delta\lambda=3.2$\,\AA ~or
$\Delta v = 220{\rm\,km\,s^{-1}}$. On the red arm we used the
$1200{\rm\,lines\,mm^{-1}}$ grating, blazed at 8000\,\AA, that covers
7600--8900\,\AA\ at $\Delta\lambda = 2.1$\,\AA\ or $\Delta v =
75{\rm\,km\,s^{-1}}$.  The blue-side detector is a mosaic of two
Marconi CCDs with $4096\times 4096$ pixels 15\,$\mu$m on the side,
which we read out binned by two in the dispersion direction.  The
red-side detector is a Tektronic CCD with $2048\times 2048$ pixels
24\,$\rm \mu$m on the side, which we read-out unbinned.

The spectra were extracted and calibrated as above.  Here, on the blue
arm we replaced the poorly exposed part of the flat fields shortward
of 4000\,\AA\ with unity and normalized the rest using a third degree
polynomial.  On the red side we normalized the flat field using a
bi-linear fit.  Wavelength calibration was done using arc spectra and
sky lines. On the blue arm we used the well exposed arc frames taken
at the beginning of the night to establish an overall solution that
had rms residuals of 0.16\,\AA\ for 22 lines fitted with a
third-degree polynomial and then calculated offsets using the less
well-exposed arc frames taken throughout the night.  For the red arm
we used the well exposed arc-frames taken interspersed with the
science exposures.  Here, we corrected for offsets by shifting the
bright $\rm{OH}$ and $\rm{O}_2$ lines at 8344.602, 8430.174 and
8827.096\,\AA\ to laboratory values.  Flux calibration was again done
as above; we found that the solution was consistent with that obtained
from Gemini (see also below).

\subsection{Photometry}
On the night of 2008 February 28, images of the field containing
PSR\,J1738+0333 were acquired for us with the 4.1\,m Southern
Astrophysical Research Telescope (SOAR) at Cerro Pach\'on, Chile,
using the Goodman High Throughput Spectrograph \citep{soar}, with its
Fairchild $4096\times 4096$ CCD and $B$ and $V$ filters (with
throughputs on the Kron-Cousins photometric system).  The instrument
has a plate scale of $0\farcs15{\rm\,pix^{-1}}$ and a usable field of
view of $5\farcm0$. During the run, the sky was photometric and the
seeing as determined from the images was $\sim\!1\farcs8$.  Two 300\,s
images each in $V$ and $B$ were obtained.  Of these, however, the
first had reduced count rates for all stars and a distorted
point-spread function, possibly because the telescope and instrument
had not yet settled when the exposure was started; we have not used
that image.  For calibration, sets of 30\,s $B$ and $V$ images of the
photometric standard field PG\,1633+099 were acquired both before and
after the science frames.

Following standard prescriptions, individual frames were
bias-corrected and flat-fielded using twilight flats. Hot pixels and
cosmic rays were replaced by a median over their neighbours.  The
instrumental fluxes were measured inside $3\farcs 6$ radii and then
corrected to a radius of $7''$ using measurements of bright
isolated stars.  For the calibration, we used 5 standard stars with a
range of $B-V$ colors in the PG\,1633+099 field \citep{ste90}.
Measured magnitudes were compared to their catalogued counterparts to
derive zero-points and colour terms.  Both calibration sets yielded
similar results.  Small differences in airmass were corrected using
standard values for La Silla.  The root-mean-square residuals of the
zero points in both bands were $\sim\!0.01$\,mag.  We find that the
optical counterpart of the WD has $V=21.30(5)$ and
$B=[21.70(7),21.73(7)]$, where the two measurements in $B$ are for the
two exposures, and where for the errors, we combined in quadrature the
measurement and zero-point uncertainties.  For the comparison star, we
measure $V=18.00(1)$ and $B=[18.73(2),18.75(3)]$.  Since the $B$
magnitudes are consistent, we use the averages below.

We verified our calibration in several ways.  First, we integrated our
flux-calibrated spectra over the B-band filter curve of \cite{b90}.
For the comparison star, using the wide-slit spectra, we find
$B'=18.81$ for the Gemini spectrum and $B=18.71$ for the Keck
spectrum.  For the white dwarf, we find $B'=21.69$ for the averaged
Gemini white dwarf spectra, and $B=21.70$ for the single narrow-slit
Keck spectra.  Here, we label the Gemini magnitudes as $B'$, since the
GMOS spectra do not fully cover the Bessell-B bandpass, which will
introduce color terms.

Second, we tried to calibrate the $g'$-band GMOS acquisition images,
by calibrating relative to our velocity standard, WD~1743$-$132, which
has $V=14.290$, $B-V=0.300$ \citep{wdvel}, and
thus, using the relations of \cite{sloan}, $g'=V+0.56(B-V)-0.12=14.34$.  We find
$g'=18.23$ for the comparison star and, using the average magnitude
difference $\Delta g'=3.091(17)$ between the WD and the comparison,
we infer $g'=21.32$ for the WD (here, the uncertainty will be dominated
by systematics, but should be $\la\!0.05\,$mag).  These numbers are
consistent with the $g'=18.39(7)$ and 21.42(7) expected from our SOAR
photometry.

Looking at individual acquisition frames, the scatter of the magnitude
difference was $\sim\!0.05\,$mag, somewhat larger than expected based
on measurement noise, though with no obvious correlation with orbital
phase.  We find somewhat smaller scatter from convolving individual
flux calibrated WD and comparison spectra with the Bessell $B$-band,
and using those to determine differences (see Table\,\ref{table:1}).
Ignoring the two points from our worst night (2006 May 26), the
root-mean-square scatter is 0.032\,mag.  Since no obvious phase
dependence is found, this places a limit on the irradiation of the WD
atmosphere from the pulsar.  However, the limit is too weak to be
useful: Assuming a spin-down luminosity of $L_{\rm{PSR}} = dE / dt =
-4\pi^2 I \dot{P} / P^3 \sim 4.8 \times 10^{33}$\,ergs\,s$^{-1}$ and defining an
irradiation temperature $T_{\rm irr}=(L_{\rm PSR}/4\pi
a^2\sigma)^{1/4}\simeq 3800{\rm\,K}$ (where from Table~\ref{table:2},
we inferred $a\simeq1.8\times10^{11}{\rm\,cm}$), the expected orbital
modulation is only $\Delta L / L\simeq [\pi R_{\rm
 WD}^2(L_{\rm{PSR}}/4\pi a^2)/L_{\rm WD}]\sin i \simeq [T_{\rm irr}^4/4T_{\rm
 WD}^4]\sin i\la 4\times 10^{-3}$.

\section{Results}\label{section:4}
\subsection{Radial velocities} 
Radial velocities of the WD, the comparison and the velocity standard
were extracted by fitting their spectra with templates using the
method discussed in \citet{bvk+06}.  For the comparison, we first
classified it using the on-line atlas by
R.~O.~Gray\footnote{http://nedwww.ipac.caltech.edu/level5/Gray/Graycontent}.
We find that its spectrum resembles that of a G0V star, with an
uncertainty of about 1 subtype.  Comparing with various spectra from
the UVESPOP\footnote{http://www.sc.eso.org/santiago/uvespop/DATA}
library of high resolution spectra \citep{uves}, we find the best fit
for the G1V star HD~20807 (where, to match the resolution of the
observations, we convolve the UVESPOP spectra with a Gaussian with
FWHM equal to that of the seeing, truncated at the slit width).  We
fitted this template to each spectrum for a range of velocities, from
$-600$ to $600{\rm\,km\,s^{-1}}$ with a step size of
$5{\rm\,km\,s^{-1}}$.  We corrected for the $11.5{\rm\,km\,s^{-1}}$
barycentric velocity of HD\,20807 after the fact.

Similarly, the WD spectra were compared to an appropriate DA model
atmosphere.  The latter was determined iteratively, where we first
fitted a high S/N single spectrum with a grid of model atmospheres
created by one of us (D.~Koester, see next section), then used the
best fit solution to shift the spectra and average them at zero
velocity, and finally fitted the average again to determine the best
template.  For WD\,1743$-$132 we fitted the single spectrum with the
grid and determined all parameters simultaneously.

For all above fits, we multiplied the templates with a 3$^{\rm{rd}}$\,degree
polynomial to account for the normalization and possible variations
with wavelength (see \S4.5 for details).  Our best fits gave typical reduced $\chi^{2}$ values
of $\chi^{2}_{\rm red,min}\sim1.2$, 2.2 and 1.6 for the WD, the
comparison star, and the velocity standard, respectively.  Best-fit
velocities were determined by fitting a parabola to the $\chi^{2}$
values to within $60{\rm\,km\,s^{-1}}$ of minimum, with uncertainties
taken to be the difference in velocity over which $\chi^2$ increased
by $\chi^2_{\rm red,min}$ (thus effectively increasing our
uncertainties to account for the fact that $\chi^2_{\rm red,min}$ did
not equal unity).

For the Keck spectra, we proceeded similarly. Here on the red side, we
could not use the UVES spectrum due to a gap over the  Ca\,II
triplet, and hence we used instead a $T_{\rm eff}= 6000$\,K, $\log
g=4.5$\,dex model by \citet{zcm04}.  As we trust the absolute wavelength
calibration of this observation most (being calibrated relative to
telluric emission lines), we use this estimate of the velocity below
to transform all velocities to the barycentric reference frame.

\subsection{Radial-velocity orbit and mass ratio}
In Table\,\ref{table:1} we list the measured radial velocities for all
targets, with barycentric corrections applied.  For determining the
orbit, we folded the barycentric velocities using the ephemeris in
Table\,\ref{table:2} and fitted for a circular orbit keeping the
orbital period and time of accenting node passage fixed to the timing values. The fit
gave a radial velocity semi-amplitude of $K_{\rm obs} = 165 \pm
7{\rm\,km\,s^{-1}}$ and a systemic radial velocity of $\gamma = -50
\pm 4{\rm\,km\,s^{-1}}$ with $\chi^2_{\rm{red}}=1.55$ for 16 degrees of
freedom.

The radial velocity of the comparison star in the Gemini dataset
varied as much as $55{\rm\,km\,s^{-1}}$ which is considerably higher
than the uncertainties of individual points.  We found no evidence for
binarity and thus we attribute the large scatter to systematics,
likely induced by slit positioning errors and differential atmospheric
diffraction.  For that reason, we chose to use velocities relative to
the comparison star, $\Delta v$. 
This choice relies on the assumption that both the WD and the
comparison star are affected by the same systematics.  
This should be correct to first order, but given the relatively large 
separation of the two stars on the slit, their different distances from the 
centre of rotation of the instrument, and their different colours, 
small second-order differences may remain.  
Even if any are present however, they should 
not be correlated with orbital phase 
(since our measurements are taken on many different nights), 
and thus be taken into account automatically by our rescaling
of the measurement errors such that reduced $\chi^2$ equals unity.

\begin{figure}
\includegraphics[width=\hsize]{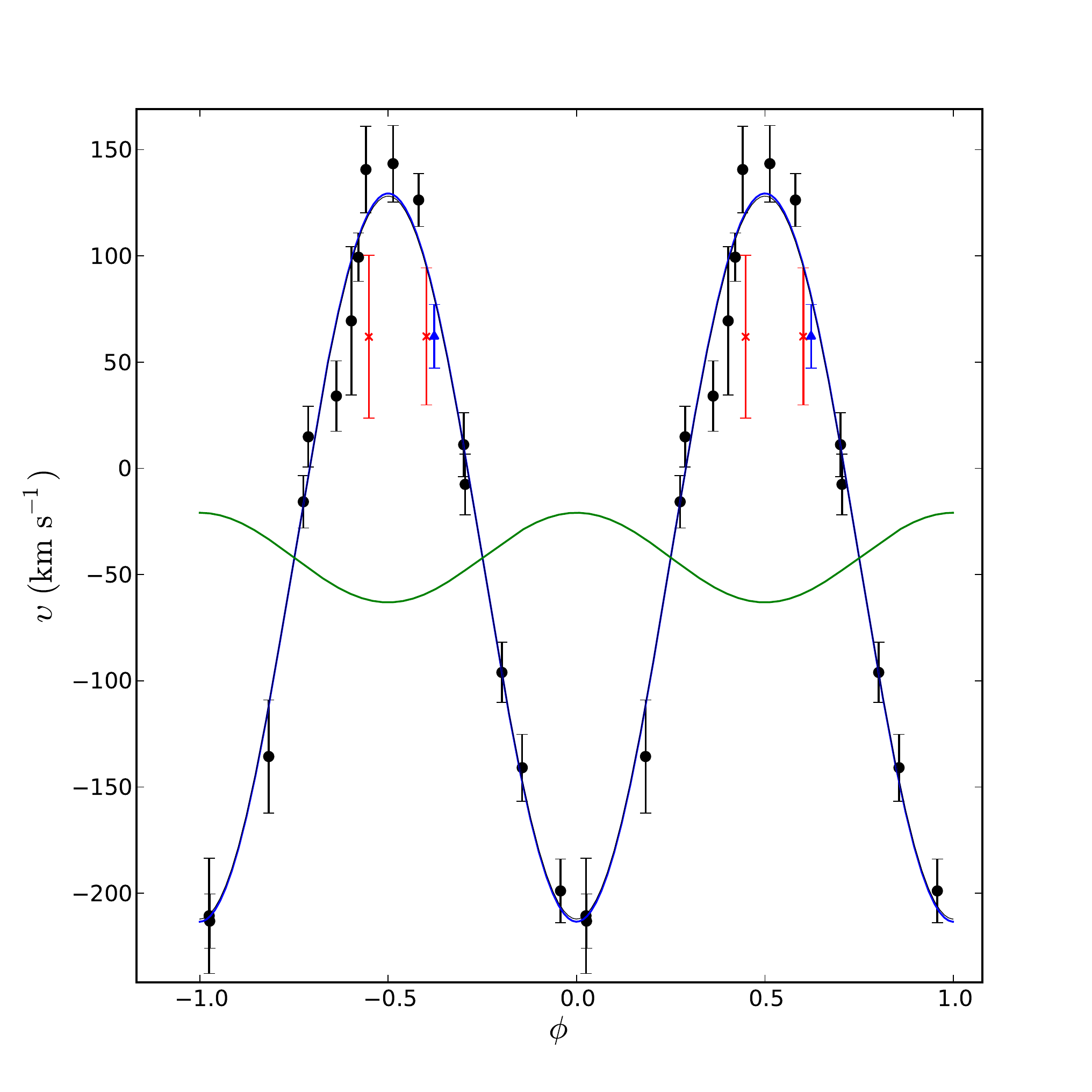}
\caption{Radial velocity measurements of the companion to
 PSR\,J1738+0333 as a function of the orbital phase.  Filled black
 circles depict the points used to fit the orbit and the blue line
 the best-fit solution. Red crosses indicate two outliers that we
 excluded and the black line the best-fit solution with these points
 included. The latter agrees well and is almost indistinguishable.
 The blue triangle shows the Keck point.  The green line
 depicts the velocity of the pulsar as inferred from radio
 timing. All velocities are relative to the comparison star, but
 corrected for its estimated $61{\rm\,km\,s^{-1}}$ barycentric radial
 velocity.  All error bars represent $1\sigma$ uncertainties. The orbit is depicted
 two times for clarity.}
\label{figure:orbit}
\end{figure}

After subtracting the velocity of the comparison star, we obtain
$K_{\rm obs}=166\pm6{\rm\,km\,s^{-1}}$,
$\Delta\gamma=-101\pm4{\rm\,km\,s^{-1}}$ with $\chi^{2}_{\rm
 red}=1.07$. This orbit is shown in Fig.\,\ref{figure:orbit}.  This
fit has two outliers, which both are from spectra taken in the night
with the worst condition (they are also outliers in the relative flux
between the WD and the comparison star; see Table~\ref{table:1}).
Excluding these, we find $K_{\rm obs}=167\pm5{\rm\,km\,s^{-1}}$ and
$\Delta\gamma=-103\pm3{\rm\,km\,s^{-1}}$ with $\chi^{2}_{\rm
 red}=0.93$ for 14 degrees of freedom.  We will use these latter
values as our best estimates, but note that all fits gave consistent
results, so our inferences do not depend on this choice.

Because the exposure time is a significant fraction of the orbit
($t_{\rm exp}\simeq0.12\,P_{\rm b}$), the observed semi-amplitude is
affected by velocity smearing.  This reduces the measured amplitude by
a factor $\sin(\pi t_{\rm exp}/P_{\rm b})/(\pi t_{\rm exp}/P_{\rm
b})=0.976$.  Thus, the true radial-velocity amplitude is $K_{\rm
WD}=171\pm5{\rm\,km\,s^{-1}}$.

Likewise, the semi-amplitude of the pulsar's projected radial velocity
is $K_{\rm {PSR}}=2\pi c x/P_{\rm b}=21.103059(2){\rm\,km\,s^{-1}}$, where
$x$ is the projected semi-major axis of the pulsar orbit.  Based on
the two values calculated above we derive a mass ratio of $q =
K_{\rm WD}/K_{\rm PSR}=8.1\pm0.2$.

\subsection{Systemic velocity}

The systemic velocity $\Delta\gamma$ derived above is relative to the
comparison star.  Thus, for an absolute value one needs to obtain an
estimate of the true velocity of the latter.  From the Gemini spectra
we derived an average value of $64\pm5{\rm\,km\,s^{-1}}$.  As
discussed above, the individual velocities have a large scatter and
one may thus worry about systematics.  It seems, that these are of the
order of $15-20$\,km\,s$^{-1}$.  First, for the velocity standard
WD\,1743$-$132 we find a velocity of $-58.6\pm1{\rm\,km\,s^{-1}}$,
which is offset by 14.2\,km\,s$^{-1}$ from the catalogue value of
$-72.8{\rm\,km\,s^{-1}}$ \citep{reid}.  Second, for the comparison
star, our Keck spectrum yields $61\pm5{\rm\,km\,s^{-1}}$.  As
mentioned above, we believe the wavelength calibration is most
reliable for the Keck spectrum, so we adapt this velocity.  For
PSR\,J1738+0333, correcting for the gravitational red-shift of the
white dwarf of $3{\rm\,km\,s^{-1}}$ (using the mass and radius derived
in \S\,\ref{section:4}.6), we infer a systemic velocity of
$\gamma=-42\pm16{\rm\,km\,s^{-1}}$.

\subsection{Interstellar reddening}
We calculated the run of reddening along the line of sight using the
Galactic extinction model of \cite{dcl03}. We find that the interstellar
extinction increases smoothly to reach a maximum value of $A_V=0.56$
at 1.3\,kpc and remains constant thereafter.  This is similar to the
maximum value along this line of sight of $A_V=0.65$ inferred from the
maps of \cite{sfd98}.  Therefore, for both PSR\,J1738+0333 and the
comparison value we adopt $A_V=0.56 \pm 0.09$, 
with the uncertainty taken to be the difference between the two models.

We can now use these results to estimate the distance of the
comparison star: Adopting $M_V = 4.3$ and $(B-V)_0 = 0.57$ for a G0V
star \citep{c00} and $A_B=1.321A_V$ \citep{sfd98} we obtain a distance
of $\sim 4.3$\,kpc for both bands.
 As a sanity check for the systemic velocity
derived above, we can calculate the expected velocity of the
comparison for the photometric parallax: Assuming the Galactic
potential of \cite{kbg+08}, a distance to the Galactic center of
8.0\,kpc and a peculiar velocity of the Sun relative to the local
standard of rest of $(U,V,W) = (10.00,5.25,7.17){\rm\,km\,s^{-1}}$
\citep{c00}, we find that the local standard of rest at the position
of the comparison star moves with a speed of $\sim\!60{\rm\,km\,s^{-1}}$.
Given the uncertainties of the model and our measurements and the
possibility of peculiar motion, the latter agrees well with our
estimated value.

\subsection{Temperature and surface gravity of the WD}

\begin{figure*}
\resizebox{17cm}{!}{\includegraphics{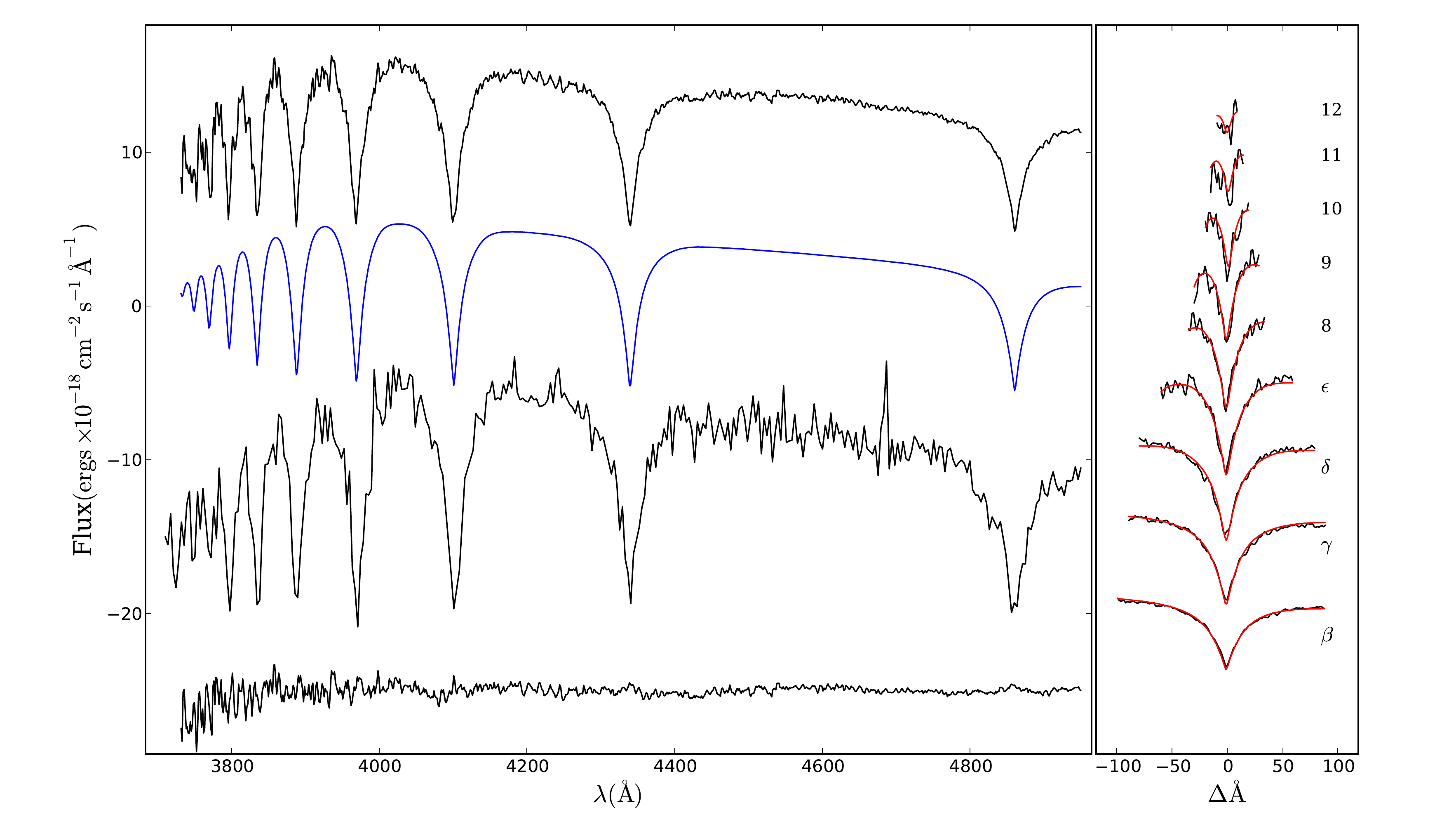}}
\caption{\textbf{Left:} From top to bottom: The zero-velocity
  flux-calibrated average spectrum of PSR\,J1738+0333 obtained with
  Gemini, the corresponding best-fit atmospheric model, the (single)
  spectrum obtained with Keck, and the residuals from the fit (see
  \S\ref{section:4}). The model and the Keck spectrum are shifted
  down by 10 and 20 units respectively. \textbf{Right:} Details of of
  the Balmer series in the average spectrum (H$\beta$ to H12, from
  bottom to top), with the best-fit model overdrawn (red
  lines).  Lines are shifted by 8 units with respect to each other.}
\label{figure:3}
\end{figure*}

The zero-velocity average spectrum (Fig.\,\ref{figure:3}) shows deep
Balmer lines up to H12, typical for a WD with a hydrogen atmosphere
and low surface gravity.

\suppressfloats

Quantitative estimates for the atmospheric parameters were obtained by
modelling the spectrum with a grid of DA model atmospheres extending
from 7000\,K to 25000\,K and $\log g = 6.00$ to $\log g = 8.00$ with
step-sizes of 100\,K and 0.1\,dex respectively.  The models used in
this work are a recent update of the grid presented in \cite{spectra}
 which incorporates the improved
treatment of pressure broadening of the absorption lines by
\citet{tb09}.

At each point of the grid that we scanned, we fitted for the
normalization with a polynomial function of the wavelength.
This was found necessary in order to account for the (up to) $\sim10$\%, slowly varying
continuum deviations, caused by in-perfect flux calibration.
Assuming our flux calibration is perfect  
 (namely, using a normalization factor that does not vary with wavelength) resulted
 in a poor fit with large scale structure in the residuals and lines systematically
 deeper than the best-fit model 
 (best-fit values: $T_{\rm{eff}}=9010 \pm 50 $\,K, $\log g = 6.81 \pm 0.12$\,dex
 with $\chi^{2}_{\rm{red}}\sim9$). Similarly underestimated lines were obtained using a
 fitting routine normally used by one of us (D.~Koester) that
 assumes a fixed slope for the continuum over the length of each line.
The former comparison 
revealed that there was also a smaller spectral range between 4400$-$4780\,\AA\ 
with features similar with the ones seen in the flat fields (see
\S\ref{section:3}), likely associated with the holographic grating
(we were alerted to this effect because it was
much more obvious for the companion of PSR J1909$-$3744; vK+12).
Fortunately, no Balmer lines are present in this region, and hence we
simply modelled the spectrum excluding this range (specifically, we fitted the
ranges 3700--4400, and 4780--4960\,\AA).  Like for our radial-velocity
fits, we accounted for the spectral resolution by convolving the
models with a truncated Gaussian.

\begin{figure*}
\centering $
\begin{array}{cc}
\resizebox{8.7cm}{!}{\includegraphics{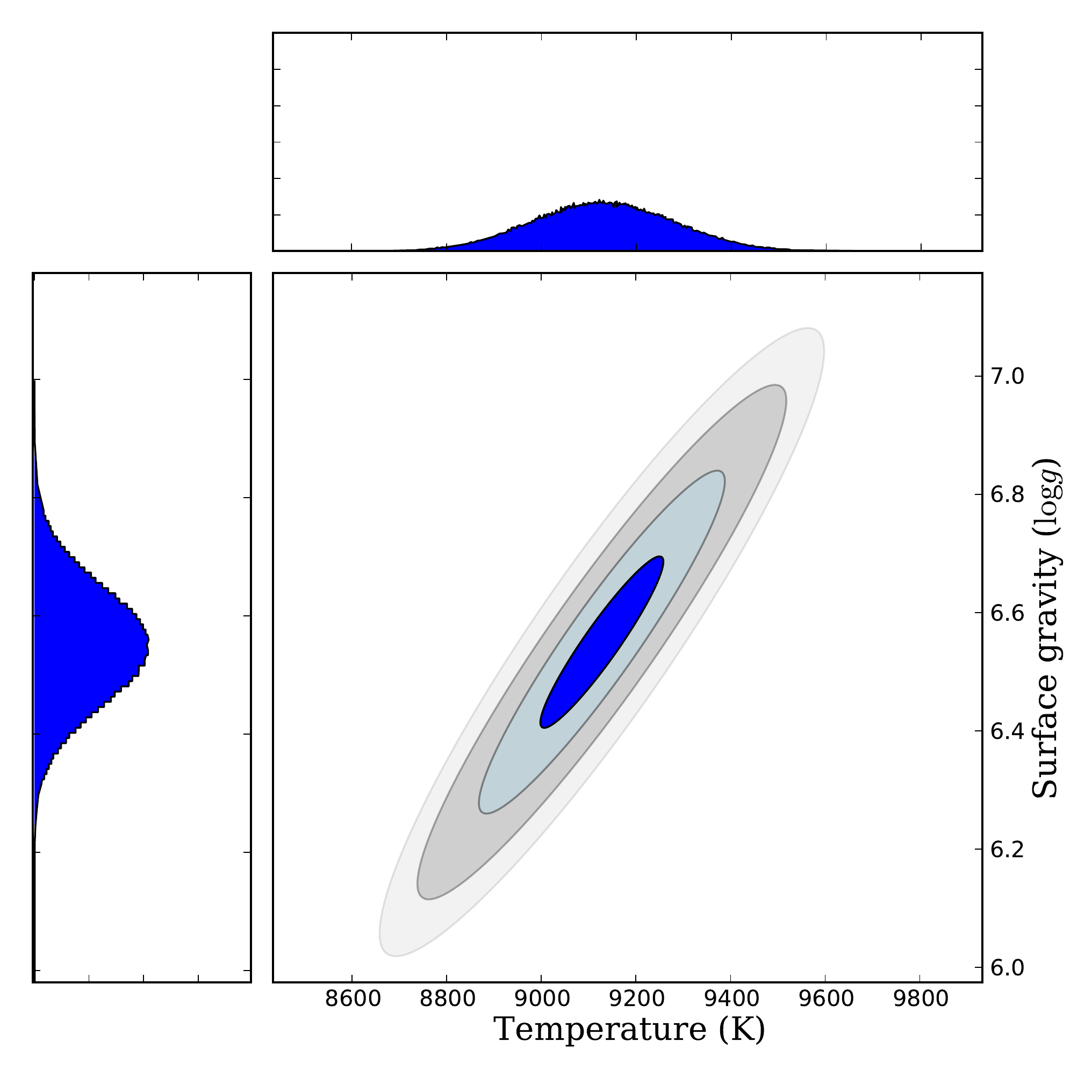}} &
\resizebox{8.7cm}{!}{\includegraphics{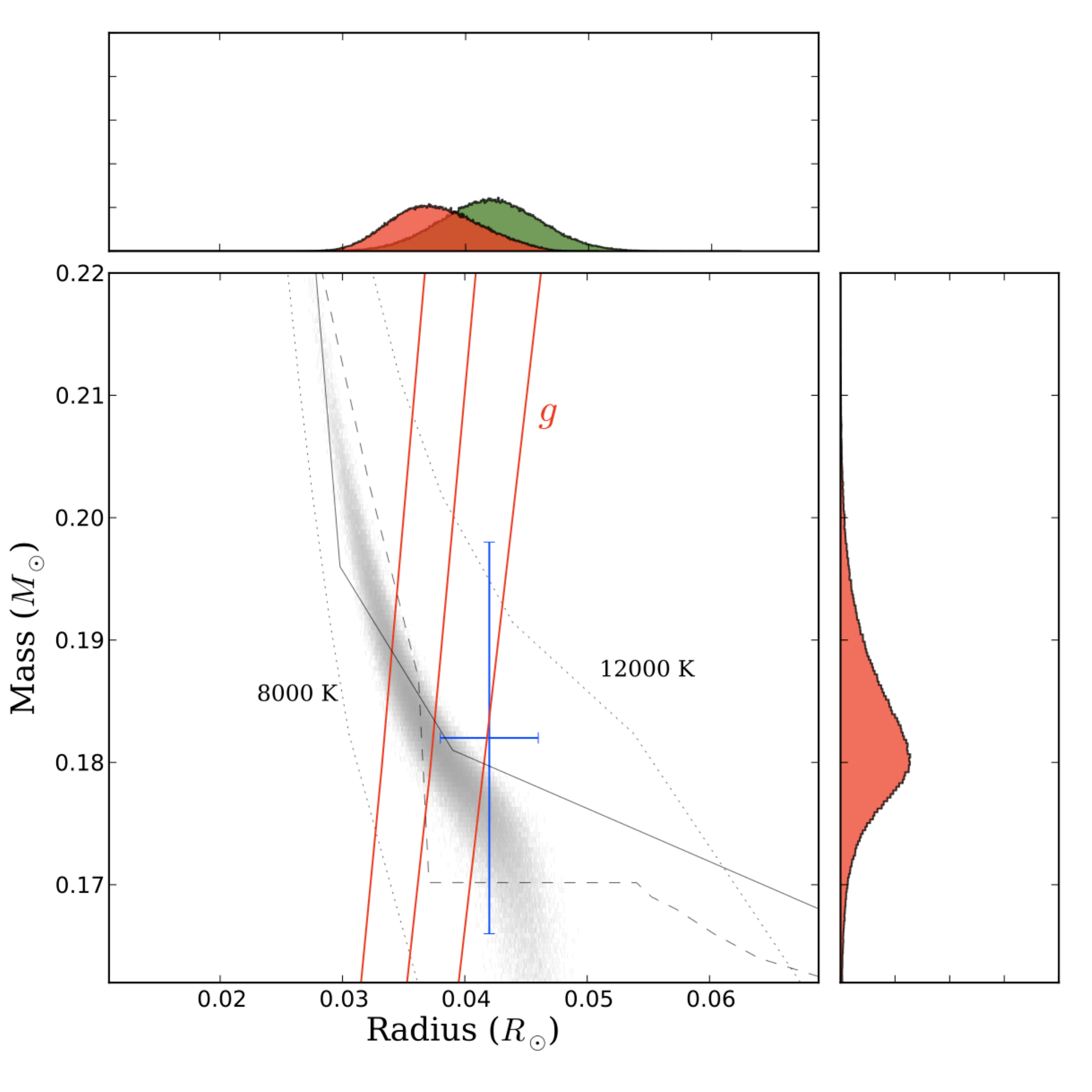}}
\end{array} $
\caption{\textbf{Left}: Constraints on the temperature and gravity of
 the white dwarf companion to PSR\,J1738+0333 inferred from our
 model-atmosphere fit, with contours at $\Delta \chi^2 = \chi^2_{\rm{red,min}}$ ,4$\chi^2_{\rm{red,min}}$, 9$\chi^2_{\rm{red,min}}$,
 and 16$\chi^2_{\rm{red,min}}$.  The horizontal and vertical sub-panels show the histograms
 of the distributions for $T_{\rm{eff}}$ and $\log g$ from our Monte
 Carlo simulation (\S\ref{section:4}.4).  \textbf{Right:} Constraints
 on the mass and radius of the WD. The shaded area depicts the distribution 
 of realizations from our Monte-Carlo simulation 
 (\S\ref{section:4}.6).  Overdrawn are: the central value and
 $1\sigma$ confidence limits of the observed surface gravity (red
 lines); the model tracks of \citet{pab00} for constant temperature (8000 and
 12000\,K; dotted lines); the mass-radius relations of \citet{sar+01} (solid) and \citet{pac07}
 (dashed) for our best-fit temperature of 9130\,K; errorbars showing
the independent constraints from photometry and radio timing.
 The horizontal and vertical panel show the inferred distributions for the WD radius and
 mass, respectively, as well as the independent photometric estimate
 for the former (in green; see \S\ref{section:4}.5). }
\label{figure:4}
\end{figure*}

Using the choices described above we obtain
$T_{\rm{eff}}=9129\pm20$\,K (implying a spectral type DA5.5) and $\log
g=6.55\pm0.07$\,dex with $\chi^2_{\rm red,min}\simeq 1.5$ (for
$\sim\!800$ points and 6 parameters).  Here, the best-fit values and
statistical uncertainties were determined by fitting the $\chi^2$
surface with a paraboloid as in \citet{bvk+06}.  We verified these
estimates using a Monte-Carlo simulation with $10^6$ iterations (see
Fig.\,\ref{figure:4}).  The results are almost identical, with the
simulation giving slightly larger uncertainties.  However, as we will
see below, the systematic uncertainties are larger.

The best-fit model is shown in Fig.\,\ref{figure:3}.  Most lines are
matched almost perfectly, but H11 and H12 are slightly underestimated.
We do not know the reason for this.  As the continuum matches very
well, it cannot be due to errors in the flux calibration (which would
be multiplicative), while most other observational issues (scattered
light, etc.) would lead to lines that have reduced rather than
increased depth.

Given the above discrepancies, as well as previous experience with
fitting model atmospheres, it is likely our uncertainties are
dominated by systematics rather than measurement errors.  We
investigated this in three ways.  First, we tried small changes in the
assumed spectral resolution (by 5 \%) and varied
the different polynomial degrees for the continuum (2nd to 4th order).
The former had only very small effect ($\sim\!20$\,K changes in $T$
and $\sim\!0.03\,$dex changes in $\log g$ ), while changing the degree
of the polynomial caused larger differences: 0.1\,dex (1.5$\sigma$)
for the surface gravity and up to 150\,K (7$\sigma$) for the
temperature.  Our central values are based on a 3rd degree
polynomial, since it gave the best fit for the higher lines.

As a second check, we obtained an independent measure of the
atmospheric parameters using the Keck spectrum.  Again using a
third-degree polynomial for the continuum, and fitting the same
wavelength regions, we find $T_{\rm eff}=9281\pm110$\,K and $\log g =
6.57 \pm 0.13$\,dex.  Here, switching between polynomials for the
continuum normalization had a slightly smaller impact on the estimated
values ($\sim\!100$\,K in $T$ and $\sim\!0.1$ in $\log g$ ).  While
the surface gravity agrees almost perfectly with the Gemini value, the
effective temperature is somewhat higher, suggesting, again, that
temperature is more sensitive to our modelling assumptions.

Finally, we fitted the individual spectra with the model atmospheres
and obtained a mean temperature of $<T>=9153\pm 38$\,K with an rms scatter of 
155\,K.

From the above, it is clear the formal uncertainty on especially the
temperature is too small, and we adopt as realistic estimates $T_{\rm
 eff}=9130\pm150\,$K and $\log g=6.55\pm0.10$\,dex.  Fortunately, the
effect of the larger temperature uncertainty on the derived masses is
small (see \S\ref{section:4}.6), because the mass-radius relation is
much more sensitive to surface gravity than to temperature.  For our
mass calculation below, we thus choose to inflate the original
$\chi^2$ map to include the systematics mentioned above but preserve
information about the covariance between parameters.

Finally we searched the average spectrum for signatures of rotational broadening.
For that we proceeded in two ways: First, we broadened a 9000\,K, $\log g = 6.5$ model
atmosphere using the analytical profile of \cite{gray} with
a limb darkening coefficient of 0.3 and scanned a grid of rotational velocities 
$0\leq v_{\rm r} \sin i \leq 1500$\,km\,s$^{-1}$
in steps of 50\,km\,s$^{-1}$. Second, we let all parameters free.
In both cases we accounted for the spectral resolution of the instrument as above.
We find the rotational broadening consistent with zero with the $1 \sigma$ upper
limit being 440 and 510\,km\,s$^{-1}$ respectively.
\subsection{White dwarf radius from photometry}

We can use the best-fit atmosphere model, the observed fluxes, and the
distance to obtain an estimate of the WD radius.  In terms of
magnitudes,
\begin{equation}
m_\lambda - 5\log(d/10{\rm\,pc})  - A_{\lambda}
                   = 43.234 -5\log (R/\rm{R_{\odot})} -2.5\log
                   F_\lambda +c_\lambda 
\end{equation}
where $m_{\lambda}$ is the apparent magnitude in band $\lambda$, the
numerical term is $-5\log(\rm{R_\odot}/10{\rm\,pc})$, $F_{\lambda}$ is the
emitted flux per unit surface area integrated over the relevant
filter, and $c_{\lambda}$ the zero-point.  Convolving the best-fit
model with the $B$ and $V$ band passes of \citet{b90} yields $F_B =
6.289 \times 10^7$\,erg\,cm$^{-2}$\,s$^{-1}$\,\AA$^{-1}$\, and $F_V =
4.353 \times 10^7$\,erg\,cm$^{-2}$\,s$^{-1}$\,\AA$^{-1}$. Here the
uncertainty due to the fit is $\sim 5 \%$ (mostly due to the
$\sim\!1.5\%$ uncertainty in temperature).  Using the zero-points of
Bessel (1990), $c_B = -20.498$ and $c_V=-21.100$, and the reddening
infered in \S\ref{section:4}.4 we obtain radii
$R=0.042\pm0.004$\,R$_{\odot}$ and $R=0.042\pm0.004$\,R$_{\odot}$ for $B$
and $V$, respectively (with the uncertainty dominated by the
uncertainty in the parallax).

\subsection{Masses of the white dwarf and the pulsar}
The mass of the WD can be estimated using a mass-radius relation
appropriate for low mass helium white dwarfs.  We use the
finite-temperature relation for low-mass WDs from \citet{pab00}, which
gave good agreement for the companion of PSR J1909$-$3744 (vK+12).

For the calculation we proceeded as follows: We sampled the inflated $\chi^2$
surface derived in \S\ref{section:4}.3 in a Monte-Carlo simulation
using $10^6$ points uniformly distributed in the $T_{\rm{eff}} - \log
g$ plane.  For each point within the expectations, we linearly
interpolated the 8000 and 12000\,K models of \cite{pab00} for WDs with
extended hydrogen envelopes to the given temperature and calculated
the mass and radius at the cross-section of the observed value (which
scales as $g = GM/R^2$) and the model.  Subsequently, we calculated
the mass of the pulsar, assuming a normal distribution for the mass
ratio with $q = 8.1 \pm 0.2$ (see \S\ref{section:4}.2).  Furthermore,
we calculated the inclination using the mass function $f_{\rm{M}}$ of the
binary ($\sin^3 i = f_M(M_{\rm{WD}} + M_{\rm{PSR}})^2 /
M^3_{\rm{WD}}$).

We show the mass distribution in Fig.\,\ref{figure:4}.  Since the
mass-radius relation is steeper towards higher masses, the companion's
mass distribution is asymmetric, with larger wings towards higher
masses. The same holds for the distribution for the radius, with
larger wings towards smaller radii.  The error on the pulsar mass is
 dominated by the uncertainties in the companion's mass
estimate.  To summarize, the values that we will be using for the rest
of this paper are: $M_{\rm{WD}} = 0.181^{+0.007, +0.017}_{-0.005,
 -0.013}$\,M$_{\odot}$, $M_{\rm{PSR}} = 1.47^{+0.07, +0.14}_{-0.06,
 -0.08 }$\,M$_{\odot}$, $R_{\rm{WD}} =
0.037^{+0.004,+0.007}_{-0.003,-0.006}$\,R$_{\odot}$ and $i =
32.6^{\rm{o} +1.0,+2.1}_{-1.0,-2.1}$.
Here, the errors separated by commas are the corresponding $68\%$ and $95\%$  
intervals spanned by the Monte-Carlo realizations.

Finally, we also derived mass estimates using two different 
sets of tracks, that gave reliable results for PSR\,J1909$-$3744 (vK+12):
The tracks of \cite{sar+01} yielded $M_{\rm{WD}} =
0.183^{+0.007,+0.011}_{-0.004,-0.005}$\,M$_{\odot}$ and $R_{\rm{WD}}
=0.037^{+0.005,+0.007}_{-0.004,-0.007}$\,R$_{\odot}$, almost identical 
to the above.
 The tracks of \cite{pac07} yielded slightly different values: $M_{\rm{WD}} =
0.175^{+0.017+,0.029}_{-0.005,-0.006}$\,M$_{\odot}$ and $R_{\rm{WD}}
=0.038^{+0.005,+0.010}_{-0.003,-0.004}$\,R$_{\odot}$. However, we note
that these models predict a cooling age much smaller than the
characteristic age of the pulsar (see next section).

\subsection{Cooling age}
We compared the absolute photometric magnitudes in $B$ and $V$ with
the theoretical cooling tracks of \citet{sar+01} for solar metallicity
progenitors to infer the cooling age of the WD.  We did this by
minimizing a $\chi^{2}$ merit function based on the sum of differences
between observed and model fluxes in both bands. The track of
\citet{sar+01} closest in mass to the companion of PSR\,J1738+0333 is
that of a 0.169\,M$_{\odot}$, for which we find $\tau_{\rm c} \sim
4.2$\,Gyr. For that age and mass, the predicted temperature and
surface gravity are $T_{\rm{eff}}\sim 8500$\,K and $\log g \sim
6.35$\,dex. For our best-fit spectroscopic estimates the same track
yields $\tau_{\rm c} \sim 2.6$\,Gyr.  Since the observed mass is
slightly heavier, its cooling age must be somewhat lower. Using the
$0.193$\,M$_{\odot}$ track, we get $\tau_{\rm{c}} \sim 600$\,Myrs. The
large difference is due to the dichotomy around $0.2$\,M$_{\odot}$
expected between WDs with thick and thin hydrogen atmospheres.  
Using the tracks of \cite{pac07}, for the mass of $0.175$\,M$_{\odot}$ inferred
using those, we again find short ages, $\tau_{\rm{c}} \sim 500$\,Myrs
from the photometry and $\tau_{\rm{c}} \sim 450$\,Myrs for the
spectroscopic parameters.

Finally, the suggested relatively large age of the system (4\,Gyr
plus 2--10\,Gyr for the progenitor to have evolved ) motivated us to
compare our observations with models for lower metallicity
progenitors. Using the 0.183\,M$_{\odot}$, $Z=0.001$ track of
\citet{sar+02} we obtain $\tau_{\rm{c}} \sim 5$\,Gyr.

The above analysis demonstrates that with the current set of observations
it is difficult to constrain the cooling age of the WD, since
this depends on both the thickness of the WD envelope and the metallicity of
its progenitor. Future, more precise constraints on the parallax and consequently
on the radius, might help to discriminate between different cases.

\subsection{3D velocity and Galactic motion}
In \S\,\ref{section:2} we computed the two components of the
transverse velocity based on the parallax and proper motion estimates
from radio timing measurements of the pulsar.  Combined with the
systemic radial velocity $\gamma = -42 \pm 16{\rm\,km\,s^{-1}}$ from
the optical observations of the white dwarf (Table~\ref{table:1}), we
have the full 3D velocity and can compute the Galactic path back in
time (like was done for PSR\,J1012+5307 by \citealt{lwj+09}). For our
calculations we have used the Galactic potential of \citet{kbg+08},
verifying our results with those of \cite{kg89} and \cite{pa90}. We
infer that the PSR\,J1738+0333 system has an eccentric orbit with a
Galacto-centric distance between 6 and 11\,kpc, and an oscillating
$Z$-motion with an amplitude of 1\,kpc and a (averaged) period of
$~125\,{\rm Myr}$. We also calculated the peculiar velocity of the
system with respect to the local standard of rest at every transition
of the Galactic plane ($Z=0$) during the last 4\,Gyr, and find that
it ranges between 70 and 160\,km\,s$^{-1}$.  We will discuss this
further in \S\ref{section:5}.1.

\section{Ramifications}\label{section:5}
\begin{figure}
\includegraphics[width=\hsize]{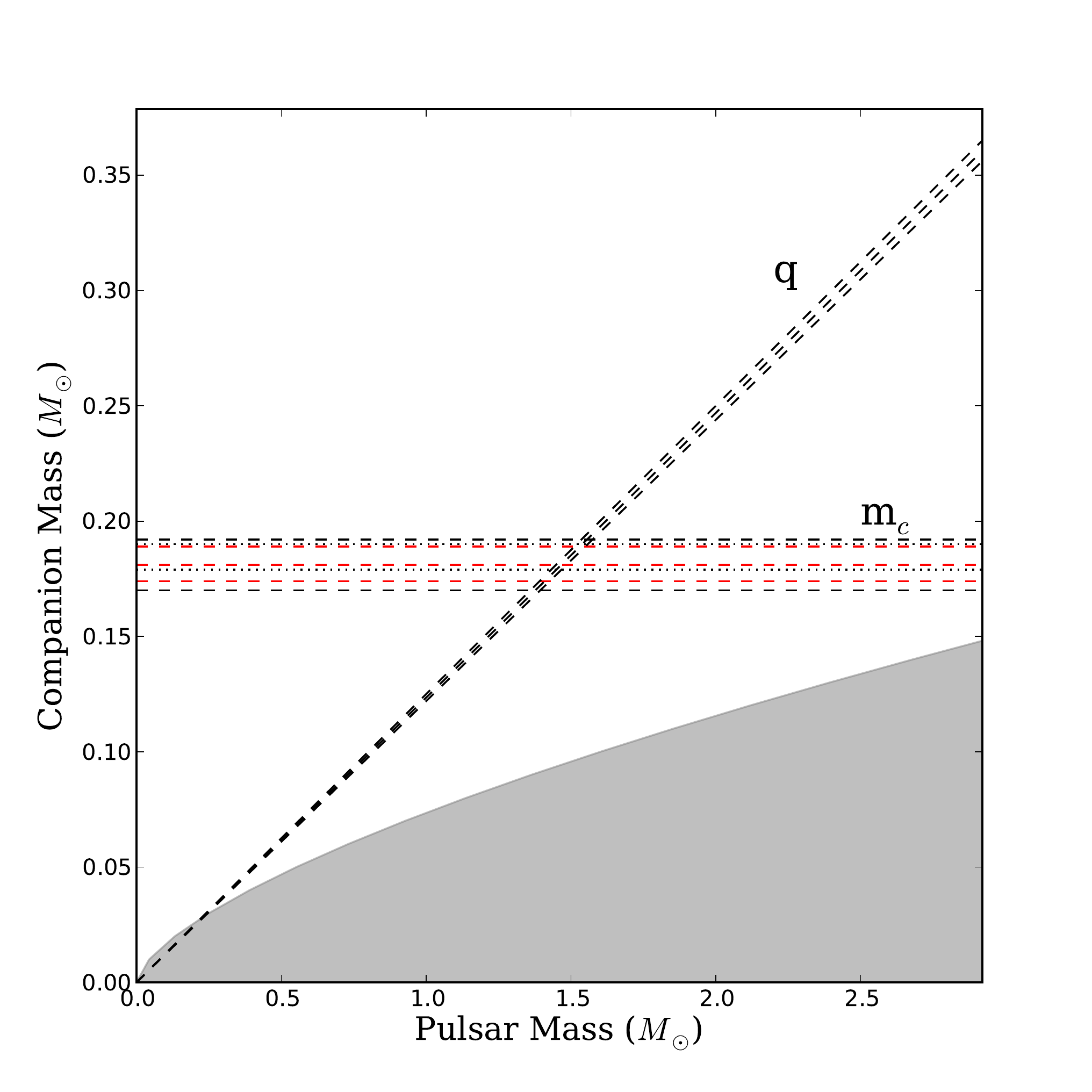}
\caption{Constrains on the masses derived from our observations.  The
  shaded region is excluded because it would require $\sin i>1$.  Our
  constraint on the mass ratio is indicated by diagonal lines, and
  that on the WD mass by horizontal lines, with the red ones giving
  the central value and $1\sigma$ uncertainties using the mass-radius
  relation of \citet{pab00}, the black dashed ones the $1\sigma$ range
  inferred using the models of \citet{pac07} and the black dotted the $1\sigma$
  limit based on \citet{sar+01}.}
\label{figure:6}
\end{figure}

In Table~\ref{table:2} we list the properties of the system derived
in previous sections and in Fig.~\ref{figure:6} we show our
constrains on the masses. In
what follows we discuss the ramifications of our work for stellar and binary astrophysics.

\begin{table}
 \caption{Properties of the PSR\,J1738+0333 system.  Numbers in
   parentheses (where given) are the formal errors of the best-fit
   model. For details on the timing analysis, including
   uncertainties not relevant here, see Paper\,II.}
\resizebox{8.5cm}{!}{
\begin{tabular}{lr}
\hline
Timing parameter & Value \\
\noalign{\smallskip}
\hline
\noalign{\smallskip}
Reference Epoch (MJD) & 54600\\ 
Time of ascending node (MJD) & 53400.297958820(6) \\
Right ascension, $\alpha$ (J2000)&17$^{\rm h}$38$^{\rm m}$53\rlap.$^{\rm s}$.965\\
Declination, $\delta$ (J2000) & $03\deg33\arcmin10\farcs866$\\ 
$\mu_{\alpha}$ (mas\,yr$^{-1})$&+7.058(5)\\
$\mu_{\delta}$ (mas\,yr$^{-1})$&+5.176(10)\\
Parallax, $\pi$ (mas)&0.67(4)\\ 
$P$ (ms)& 5.85 \\ 
$\dot{P}$ (s\,s$^{-1}$) &  2.412$\times 10^{-20}$\\  
Dispersion measure, DM (cm$^{-3}$ pc) & 33.77\\ 
Orbital period, $P_{b}$ (days)&0.35479\\ 
Projected semi-major axis, $x$ (lt-s)&0.3434\\ 
Eccentricity, $e$ &$3.5(1.1)\times 10^{-7}$\\
Mass function, $f(\rm{M_{\odot}} )$ & 0.0003455012(12) \\
\noalign{\smallskip} 
\hline
Optical parameter & Value \\
\noalign{\smallskip}
\hline
\noalign{\smallskip}
Temperature (K) \dotfill & 9130(150)\\ 
Surface gravity ($\log g$, spectroscopy) \dotfill & 6.55(10)\\
Surface gravity ($\log g$, $\dot{P}_{\rm b} + q + \pi +$photometry)\dotfill & 6.45(7)\\
Photometry, $V$-band \dotfill & 21.30(5) \\ 
Photometry, $B$-band \dotfill & 21.71(4) \\
Semi-amplitude of radial velocity, $K_{\rm{WD}}$ (km\,s$^{-1}$) \dotfill & 171(5)\\
Systemic radial velocity, $\gamma$ (km\,s$^{-1}$) \dotfill & $-$42(16)\\
Transverse velocity, $v_T$ (km\,s$^{-1}$) \dotfill & 59(6)\\
3D velocity amplitude (km\,s$^{-1}$) \dotfill & 72(17)  \\ 
Mass ratio, $q$ \dotfill & 8.1(2)\\[.8ex]
WD mass, $M_{\rm{WD}}$ (M$_{\odot}$, spectroscopy) \dotfill  & $0.181^{+0.007}_{-0.005}$\\[.8ex]
WD mass, $M_{\rm{WD}}$ (M$_{\odot}$, $q$ + $\dot{P}_{\rm{b}}$) \dotfill & $0.182 \pm 0.016$ \\[.8ex]
WD radius (Spectroscopy) (R$_{\odot}$) \dotfill & $0.037^{+0.004}_{-0.003}$ \\[.8ex]
WD radius (Photometry) (R$_{\odot}$) \dotfill & 0.042(4) \\[.8ex]
Cooling age, $\tau_{\rm{c}}$ (Gyr) \dotfill & 0.5 -- 5 \\[.8ex]
Pulsar mass, $M_{\rm{PSR}}$ (M$_{\odot}$) \dotfill & $1.47^{+0.07}_{-0.06}$ \\[.8ex]
inclination, $i$ (degrees) \dotfill & 32.6(1.0)\\
\noalign{\smallskip} 
\hline
\end{tabular}
}
\label{table:2}
\end{table}

\subsection{Kinematics}
PSR\,J1738+0333 has a velocity of $85 \pm 17$\,km\,s$^{-1}$ with
respect to the local standard of rest that co-rotates with the Galaxy
$(Z=0)$ at the distance of the pulsar.  The latter compares well with
the mean transverse velocity for the bulk of MSPs with measured proper
motions \citep[e.g. $\sim 85$\,km\,s$^{-1}$ according to][]{hllk05}.
Our semi-quantitative analysis in \S\ref{section:4}.8 shows that the
system's velocity varies as much as 150\,km\,s$^{-1}$ over the course
of its Galactic orbit.  Based on the simplified potential of
\citet{kbg+08} used herein, PSR\,J1738+0333 has a peculiar velocity
 between 70\,km\,s$^{-1}$ and
160\,km\,s$^{-1}$ when it crosses the Galactic plane $(Z=0)$.
  Thus, assuming that the system had a small peculiar
motion before the SN explosion, the systemic velocity after the formation
of the NS
 must have been in that
range.  This is consistent with a SN explosion with a small, or even
negligible kick \citep{tb96,nt95}.

\subsection{Evolutionary history}
Millisecond pulsars with low-mass helium WD companions are expected to form through mainly two
different channels depending on the initial separation of the progenitor binary \citep[e.g.][and references  therein]{t11}. The initial separation of the progenitor binary determines the evolutionary status of the 
donor star at the onset of the Roche lobe overflow (RLO):

\begin{itemize}
\item \textit{Case\,A RLO:}
 For systems with initial periods short enough to initiate mass transfer
on the main-sequence, it is expected that magnetic-braking (aided to some extent by gravitational radiation) drives the system to shorter periods,
resulting in a compact binary in an orbit which is close to being
perfectly circular (the eccentricity, $e<10^{-5}$). 
These systems were first studied in detail by \citet{ps89}.
\item \textit{Case\,B RLO:} For progenitors
with larger initial separations the mass transfer is expected to start at a
later phase, since the star fills its Roche lobe only during shell hydrogen burning,
while moving-up the red giant branch. 
In this case the orbit will diverge resulting in a wider binary.
Interestingly, for systems following this path, there are two theoretical predictions that can be verified observationally: 
The first is a correlation between the orbital period and the mass of the
WD companion which results from the unique relation between the radius of the giant donor and the mass of its core which eventually forms the WD \citep{sav87}.
The second is a correlation between the orbital eccentricity and the orbital
 period \citep{phi92} arising because the turbulent density fluctuations in the convective envelope --- of which the size
increases in more evolved stars (wider orbits) --- do not allow for a perfect tidal circularization. 

\end{itemize}

The critical period that separates diverging from converging systems
(often called bifurcation period) is expected to be $\sim 1$\,day,
however its precise value depends on the treatment of tidal
interactions and magnetic braking \citep[e.g.][]{ps89,vvp05,ml09} and is still a
subject of debate. 
The residual eccentricity in binaries with radiative donors (i.e. those binaries 
that evolve to tight converging systems) should be closer to zero compared to
binaries in wider orbits but it is
difficult to estimate by how much, as pointed out by \citet{pk94}.

With a current orbital period of 8.5\,h, PSR\,J1738+0333 is most
likely the fossil of the former case (Case~A RLO). However it is interesting to
note that our mass estimate and the non-zero eccentricity derived in
Paper\,II pass both tests for the latter case mentioned above (Case~B RLO) that
predict $m_{c} = 0.18 \pm 0.01$\,M$_{\odot}$ \citep{ts99a} and $e \sim 4\times
10^{-7}$, respectively \citep[deduced by extrapolating the][relation to
the observed period]{phi92}.  This apparent agreement seems to be
confirmed not only for PSR\,J1738+0333 but also for the other
short-period LMWD binaries with measured masses (PSR\,J1012+5307,
\citealt{lwj+09}; PSR\,J0751+1807, \citealt{nsk08}), as well as
low-mass WD companions to non-degenerate stars (e.g.,
\citealt{vrb+10,brv+11}). Since companion masses in converging
systems are not expected to follow these relations, we cannot exclude a coincidence, but the matches seem to suggest that there is a grey zone with properties from both cases -- something which should help improve our as yet rather simplified models of these systems.

\subsection{Pulsar mass and efficiency of the mass transfer}
Regardless of the evolutionary path followed, the mass transfer was
sub-Eddington \citep[e.g.][]{ts99a} and thus one would expect that a
substantial fraction of the mass leaving the donor was accreted by the
neutron star.  For PSR\,J1738+0333, this is demonstrably false: The
minimum mass of the donor star can be constrained from our WD mass
estimate to be $\ga 1$\,M$_{\odot}$ because the available time for
evolution is limited by the Hubble time (minus the cooling age of the
WD). The amount of mass lost by the donor is $M_{\rm{donor}} -
M_{\rm{WD}}$, while the amount accreted by the pulsar is $M_{\rm{PSR}}
- M_{\rm{PSR}}^{\rm{init}}$, with the last term being the birth mass
prior to accretion.  
For any realistic birth mass of the neutron star at the low end
of its ``canonical`` birth mass range
($\ga\!1.20\,M_\odot$), we find that more than 60\% of the in-falling matter must have
escaped the system (after correcting for the conversion from
baryonic mass to gravitational mass). 
This translates to an accretion efficiency of only $\varepsilon < 0.40$.
This result
confirms the findings of \cite{ts99a} who concluded that a
substantial fraction of the transferred matter in LMXBs is lost from the
system, even at sub-Eddington mass-transfer rates. 

More constraining (but less stringent) estimates are also obtained for
the 6.3~h orbital period binary, PSR\,J0751+1807 \citep{nsk08} for
which we find $\varepsilon \sim 0.1-0.3$.  

Possible mechanisms for mass ejection discussed in the literature
include propeller effects, accretion disc instabilities and direct irradiation of the
donor's atmosphere from the pulsar \citep[e.g.,][]{is75,vp96,dlhc99}. 
Alternatively, the neutron star in PSR\,J1738+0333 might
have formed via the accretion-induced collapse of a massive ONeMg WD
\citep[e.g.,][]{nk91}. If the neutron star was formed towards
the end of the mass transfer it would not have accreted much since
its birth. A possible problem with the above mechanism however,
is that it is specific to pulsars, while similarly inefficient accretion
has been found also for low-mass WDs with non-degenerate companions 
 (e.g., \citealt{brv+11}), suggesting the problem in our
 understanding is more general.
Finally, we note that even major inefficiencies in the mass accretion
process do not pose a problem for the recycling scenario: the accreted
mass needed to spin-up a pulsar to a $\sim5$\,ms period is only of the
order of 0.05\,M$_{\odot}$.

\section{Conclusions}\label{section:6}
The main result of this work is the determination of the component
masses of the PSR J1738+0333 system and adds to the three previously
known MSP-LMWD binaries with spectroscopic information
(PSR\,J1012+5307, \cite{vbk96} and \cite{cgk98}; PSR\,J1911$-$5958A,
\cite{bvk+06}; PSR\,J1909$-$3744, vK+12).

Our mass estimates are derived independently of any strong field
effects and thus transform the PSR J1738+0333 system into a
gravitational laboratory, which -- due to its short orbital period,
gravitationally asymmetric nature, and timing stability -- provides
the opportunity to test the radiative properties of a wide range of
alternatives to GR (see Paper II for details).

Based on our measurements of the component masses, GR predicts an
orbital decay of $\dot{P}_{\rm{b}} = -2.77^{+0.15}_{-0.19} \times
10^{-14}$.  While the actual $\dot{P}_{\rm{b}}$ inferred observationally
is still less precise than this prediction, it will eventually provide
a precise test for the input physics of atmospheric and evolutionary
models. 
Assuming the validity of GR, one can confront the spectroscopic
WD mass estimate  implied by the mass ratio and intrinsic
orbital decay of the system and thus test the assumptions for stellar
astrophysics and WD composition that were used to model the evolution of the WD.
Additionally, this mass estimate, combined with parallax and absolute photometry
constrains independently the surface gravity of the WD.
The current estimates on these parameters imply a surface gravity of 
$\log g = 6.45 \pm 0.07$\,dex. While this is formally more 
accurate than our spectroscopic constraint, it might still be dominated
by systematics on the distance, arising from correlations between
the parallax and DM variations (see Paper\,II for details).

Finally, the interpretation of the mass estimates within the context
of our current understanding for binary evolution implies that a
significant fraction of the accreted material during the LMXB phase is
ejected by the system. The discovery and study of more similar systems
in the future will allow further tests of this result.



\section*{Acknowledgements}
We thank the Gemini staff for their expert execution of our
observations, Rachel Rosen and Susan Thompson for obtaining the SOAR
data, and the MPIfR pulsar group for discussions.  The results of this
paper are mostly based on observations obtained at the Gemini
Observatory, which is operated by the Association of Universities for
Research in Astronomy (AURA) under a cooperative agreement with the
NSF on behalf of the Gemini partnership: the National Science
Foundation (United States), the Science and Technology Facilities
Council (United Kingdom), the National Research Council (Canada),
CONICYT (Chile), the Australian Research Council (Australia), CNPq
(Brazil) and CONICET (Argentina).  Some of the data presented herein
were obtained at the W.M. Keck Observatory, which is operated as a
scientific partnership among the California Institute of Technology,
the University of California and the National Aeronautics and Space
Administration.  The Observatory was made possible by the generous
financial support of the W.M. Keck Foundation.  The SOAR Telescope is
a joint project of: Conselho Nacional de Pesquisas Cientificas e
Tecnologicas CNPq-Brazil, The University of North Carolina at Chapel
Hill, Michigan State University, and the National Optical Astronomy
Observatory. Balmer/Lyman lines in the models were calculated with the modified 
Stark broadening profiles of Tremblay \& Bergeron (2009),
kindly made available by the authors. 
We thank J. Panei and collaborators for sharing their 2007 models. 
P.F. gratefully acknowledges
the financial support by the European Research Council for the ERC
Starting Grant BEACON under contract no. 279702
This research made extensive use of NASA's ADS and CDS's
SIMBAD services.

\bibliographystyle{mn2e}
\bibliography{psrrefs.bib}
\end{document}